\documentstyle[12pt]{article}
\begin{document}
\begin{flushright}
IP/BBSR/93-82\\
hepth-9312067\\
\end{flushright}
\begin{center}
{\bf Anyon in an External Electromagnetic Field:\\
Hamiltonian and Lagrangian Formulations}
\end{center}

So far we do not know any formulation of anyons as relativistic
free particles therefore even a small step
in this direction has considerable significance.
Recently, Chaichian, Felipe,
and Martinez [1] (hereafter referred to as I) have proposed
a model for free relativistic particles with arbitrary spin.
In this note we wish to point out that in I the Dirac
quantization procedure is not properly followed as a result
of which the resulting quantum theory is different from what one
will get by a straightforward application of Dirac's formulation.
For example at quantum level the model in I has noncommutative
geometry, i.e., $[x_\mu, x_\nu] \neq 0$ while a correct treatment
would give $[x_\mu, x_\nu] = 0$. Moreover, in I, the
assertion about $\alpha$ being an arbitrary constant
in $S_\mu = -\alpha p_\mu /{\sqrt{p^2}} $ is not correct
and consequently one can not say
whether the model describes anyons.

The Lagrangian chosen in I is
$L = m({\dot x}{\dot n})/{\sqrt{{\dot n}^2}}$ where the vector $n_\mu$
which  describes the spin degree of freedom of the particle is taken to be
spacelike and hence  there is a constraint $n^2 + 1 = 0.$ This constraint
should have been included in the Lagrangian or the Hamiltonian as is
appropriate for a constrained system. Consequently, in I
the authors were forced to take the Poisson bracket(P.B.) of $n_\mu$ with its
canonically conjugate momentum to be $\{n_\mu, p^{(n)}_\nu\}
= -(g_{\mu\nu} + n_\mu n_\nu)$ while in the canonical case
the second term on the right hand side
should be absent.

The canonical description starts by incorporating the
constraint in the Lagrangian, i.e., by taking
\begin{equation}
L = m({\dot x}{\dot n})/{\sqrt{{\dot n}^2}}
+ \lambda (n^2 + 1).\\
\end{equation}
The P.B. are all canonical and the canonical Hamiltonian
is $H = -\lambda (n^2 + 1) $. Following Dirac's method we obtain three
primary constraints
\begin{equation}
\phi_{1\mu} \equiv p_\mu -
{{m\dot n_\mu}\over {\sqrt{\dot n^2}}}\approx 0;\\\nonumber
\phi_{2\mu} \equiv p^{(n)}_\mu - {m\over {\sqrt{\dot n^2}}}
({\dot x}_\mu - {{(\dot x\dot n){\dot n}_\mu}\over {{\dot n}^2}})\approx 0;
\phi_3 \equiv p^{(\lambda)} \approx 0.\label{pricon}
\end{equation}
After incorporation of these constraints the Hamiltonian becomes
\begin{equation}
H^* = \lambda_1^\mu \phi_{1\mu} +\lambda_2^\mu \phi_{2\mu}
+\lambda_3\phi_3 -\lambda(n^2+1).\label{hstar}
\end{equation}
At this point also there is departure from  Dirac's procedure in I.
More specifically the constraints implemented
in I are not the linear combinations
of the constraints obtained by Dirac's prescription.

Now we have to  take care of the consistency conditions following
from the fact that the constraints should be conserved in
time, i.e., their P.B. with $H^*$ should vanish. We find that
$\phi_{1\mu}$ is identically conserved. P.B. of $\phi_{2\mu}$
with $H^*$ gives $\lambda n_\mu = 0$. In Dirac's classification
of consistency conditions this equation would
determine the Lagrange multiplier $\lambda = 0$.
The consistency condition corresponding to
conservation of $\phi_3$ gives $n^2 +1 \approx 0$. This has to
be incorporated as a secondary constraint
\begin{equation}
\phi_4 \equiv n^2 + 1 \approx 0.\label{phi4}
\end{equation}

Since one does not differentiate between primary and secondary
constraints therefore the complete set of constraints is
$\{{\phi_1}_\mu, {\phi_2}_\mu, \phi_3, \phi_4\}$ and the total
Hamiltonian is
\begin{equation}
H^T =  \lambda_1^\mu \phi_{1\mu} +\lambda_2^\mu \phi_{2\mu}
+\lambda_3\phi_3 + \lambda_4\phi_4.\label{htotal}
\end{equation}
But one still has to  incorporate the consistency conditions. One  finds
that $\phi_{2\mu}$ and $\phi_4$ are the only constraints giving
rise to nontrivial consistency conditions which are
$\lambda_2^\mu n_\mu = 0$ and $\lambda_4n_\mu = 0$. These equations
fix $\lambda_{2\mu}$ and $\lambda_4$ to be zero but we need not
worry about these solutions because the corresponding constraints
turn out to be second class and thus are identically equal to
zero in quantum theory. Thus one does not get any more secondary
constraints.

At this point we would like to remark that in I the consistency
conditions have not been fully
taken into account but fortunately they only
determine some of the Lagrange multipliers and do not give rise
to new constraints.

Now by computing the various P.B.'s  of the constraints
one can classify them as first (commuting) and second (rest)
class. One finds that ${\phi_1}_\mu$ and $\phi_3$ are first
class constraints while the remaining constraints
are second class because
\begin{equation}
\{\phi_{2_\mu}, \phi_4\} = 2n_\mu.\label{2ndclass}
\end{equation}
The Dirac matrix can now be defined as ${(C_\mu)}_{12}
=- {(C_\mu)}_{21} =\{\phi_{2\mu}, \phi_4\} = 2n_\mu $
and ${(C_\mu)}_{11} = {(C_\mu)}_{22} = 0$. Its inverse with
${(C_\mu)}^{12}=- {(C_\mu)}^{21} = n_\mu/2 $
and ${(C_\mu)}^{11} = {(C_\mu)}^{22} = 0$ can be used to define the
Dirac bracket of any two quantities $A$ and $B$ as
\begin{equation}
[A, B]_D = \{A, B\} -\{A, \phi_{2\mu}\}{(C^\mu)}^{12}\{\phi_4, B\}
-\{A, \phi_4\}{(C^\mu)}^{21}\{\phi_{2\mu}, B\}.\label{Dbracket}
\end{equation}
These brackets can be used to pass to quantum theory. One finds
that unlike the claim in I, the geometry is commutative, i.e.,
$[x_\mu, x_\nu] = 0.$ Moreover, the commutator of $n_\mu$
with its canonically  conjugate momentum does not depend on
the momentum $p_\mu$, i.e., one gets $[n_\mu, p_\nu^{(n)}]
= -i(g_{\mu\nu} + n_\mu n_\nu)$. Thus the actual quantum
theory is different from what is found in I.
One also notices that since $\phi_4$ is a second class
constraint therefore eqn(4) is strongly zero in the quantum
theory and thus we recover the spacelike condition for $n_\mu$.

Finally, in I, it is asserted  that $\alpha$
occurring in $S_\mu
= -\alpha p_\mu/\sqrt{p^2}$ is an arbitrary
constant. But if this is so then
one would get the wrong result $[S_\mu, S_\nu] = 0$ since ${p_\mu}'$s
commute. Therefore one concludes that
$\alpha$ can not be an arbitrary
constant but has to be an operator. In particular it
should depend on $x_\mu$ because its commutator with
$p_\mu$ has to be nonzero. Also $\alpha$ can not be
completely arbitrary because of the algebra
$[S_\mu, S_\nu] = i\epsilon_{\mu\nu\lambda} S^{\lambda}$.
Therefore it is not clear if the model in I describes
a free relativistic particle of fractional spin in 2+1
dimensions.

\vspace*{.5cm}
\begin{flushleft}
Abbas Ali and Avinash Khare\\
Institute of Physics, Bhubaneswar-751005, India\\
abbas, khare@iopb.ernet.in\end{flushleft}
\begin{flushleft}
PACS Numbers: 11.10.Ef
\end{flushleft}
\begin{flushleft}
[1] M. Chaichian, R.G. Felipe, D.L. Martinez,
Phys. Rev. Lett. {\bf 71}, 3405(1993).
\end{flushleft}

\end{document}